\documentclass[twocolumn,superscriptaddress,showpacs,preprintnumbers,amsmath,amssymb,floatfix]{revtex4-1}
\usepackage{graphicx}
\usepackage{dcolumn}
\usepackage[latin1]{inputenc}

\begin{document}


\title{The plastic flow of polycrystalline solids}

\author{Miguel Lagos}
\email{mlagos7@gmail.com}
\affiliation{Departamento de F{\'\i}sica, Facultad de Ciencias,
Universidad de Chile, Casilla 653, Santiago, Chile.}

\date{\today}

\begin{abstract}
A polycrystalline solid is modelled as an ensemble of random irregular
polyhedra filling the entire space occupied by the solid body, leaving no
voids or flaws between them. Adjacent grains can slide with a relative
velocity proportional to the local shear stress resolved in the plane
common to the two sliding grains, provided it exceeds a threshold. The
local forces associated to the continuous grain shape accommodation for
preserving matter continuity are assumed much weaker. The model can be
solved analytically and for overcritical conditions gives two regimes of
deformation, plastic and superplastic. The plastic regime, from yield to
fracture, is dealt with. Applications to nickel superalloys and stainless
steels give impressive agreement with experiment. Most work of the last
century relies on postulating pre--existent cracks and voids to explain
plastic deformation and fracture. The present model gives much better
results.
\end{abstract}

\maketitle

Fracture in solid materials has been dealt with for hundred years with
Griffith's theory \cite{Griffith}, which posits that solid materials have
pre--existent cracks that concentrate tension and are able to evolve when
subjected to stress. Griffith's original work, neatly explaining why glass
tensile strength is about 1/100 of what is expected from the breaking of
the atomic bonds, has influenced a century of investigations. In this
scheme a  crack in a brittle medium grows  when the energy released by
its extension is greater than, or equal to, the energy demanded to form
a new free surface. Fracture mechanics goes beyond brittle fracture
incorporating dissipative terms accounting for energy losses at the
evolving crack tips. Griffith noted that the strength of thin enough
rods of glass increases for decreasing width, and prepared rods or
fibers of glass and quartz fine enough to have tensile strengths of
about $10^6$ psi, which is a figure close to the strength one would
expect from the inter--atomic bindings. These high strengths stayed 
stable for a few hours, and then fall to much smaller constant values
which depend on the diameter of the fiber. The final strengths, about
hundred times smaller, were observed to be comparable in glass and metal
wires. Griffith interpreted that the finest glass fiber diameters were
smaller than the equilibrium mean pre--existent crack sizes, and then the
fiber have less or no flaws at all, which explains the higher strength
\cite{Griffith, Kendall}.

Much later Irwin \cite{Irwin} and Orowan \cite{Orowan} extended the idea
of the pre--existent flaws to ductile solids. Brittle character is
ascribed to the ability of cracks under stress to propagate conserving
their atomically sharp edges. Ductile solids are assumed to have a more
complex evolution, the tip of the stressed crack blunts by dissipative
processes, broadens, splits and flows, demanding increasing effort to
evolve \cite{Fineberg1,Kolvin,Svetlizky,Heizler,Noell}. Within this view,
as all bodies can be broken, pre--existent flaws are present in all
polycrystalline solids, and their properties are an interesting matter of
study \cite{Kendall}. Anyway, the complete regime of plastic flow under
Griffith's hypothesis, from yield to fracture, is not more nor less than
a detailed account of an avalanche of catastrophic failure events.

The model for the polycrystalline solid reported here is exactly opposed
to Griffith's hypothesis and most of the subsequent developments on it.
The solid is assumed as a perfect polycrystal, with no flaws at all, and
is modelled by an ensemble of random irregular polyhedra filling the
entire space occupied by the solid body, leaving no voids between them.
Under strong enough applied stress, adjacent grains can slide with a
relative velocity proportional to the local shear stress resolved in the
plane shared by the two grain boundaries, when it exceeds a finite
threshold $\tau_c$. The local forces induced by the continuous grain
reshaping in order to preserve matter continuity are assumed much weaker
than those causing sliding. The plastic flow is then controlled mainly by
the local shear forces making contiguous grain boundaries to slide
\cite{Lagos1,Lagos2,Lagos3,Lagos4}.

Grain boundary sliding is governed by the microscopic mechanism explained
in Refs.~\cite{Lagos1} and \cite{Lagos2}, and partially observed in Refs.~
\cite{Qi,Bellon} and \cite{Fukutomi}. Matter between adjacent grains,
represented by an elastic plate immersed into a different elastic medium,
buckles and corrugates at a critical shear stress operating along it. The
consequent periodic stress field exerted on the two grain surfaces causes
the release and capture of vacancies in alternate sectors of them. The
consequent streams in closed loops between the crystalline surfaces assist
their relative motion in opposite senses. Hence, a relative disorder in
the atomic scale at the grain interfaces does not reduce, but helps grain
sliding. The equations of motion for the plastic flow are derived with
no additional hypotheses, and have two families of solutions, corresponding
to plastic and superplastic deformation \cite{Lagos5,Lagos6}.

The superplastic solutions have been studied in an already published series
of  papers and successfully applied to the main superplastic materials.
The study of solids with normal ductility in this context was initiated in
a couple of recent papers \cite{Lagos3,Lagos4}, and the purpose of this
letter is to review the matching of the novel approach to normal ductility
with the phenomenology of real materials. Data on the nickel Inconel
superalloys and stainless steel AISI 316, of use in the combustion chamber
of jet engines and nuclear reactors, were chosen to compare with because
have special reliability.

Our hypothesis is that plastic deformation takes place by only the action
of the shear forces operating in the grain boundary faces, making
adjacent grains to slide. Forces involved in the consequent continuous
grain reshaping are of less importance. Additionally, the relative speed
of the two neighbouring grains is proportional to the shear stress resolved
in the plane of the common grain boundary, provided that this shear stress
is greater than a threshold $\tau_c$.

Assuming axial symmetry with respect to the $z-$axis, the stress tensor
reads 

\begin{equation}
(\sigma_{ij}) = \left( \begin{array}{ccc}
\sigma_\perp & 0 & 0 \\
0 & \sigma_\perp & 0 \\
0 & 0 & \sigma
\end{array}\right),
\label{E1}
\end{equation}

\noindent
where a non--vanishing transversal stress $\sigma_\perp$ is assumed
because it will be demonstrated that plastic strain always induces such
kind of stress, no matter the initial conditions. The hydrostatic pressure
$p=-(\sigma +2\sigma_{\perp})/3$ (the minus sign is placed just to be
consistent with pressure in a fluid) will prove to be an essential variable
in the plastic flow. 

The relative motion of two neighbouring grains is proportional to the shear
stress $\tau_{z'}=[\sigma_{x'x'}^2+\sigma_{y'y'}^2]^{1/2}$ operating in the
common boundary plane between them, provided that $\tau_{z'}>\tau_c$. Here
$(\sigma_{i'j'})$ is the stress tensor (\ref{E1}) expressed in a rotated
frame of reference with the $z'$--axis normal to the plane of the two grain
boundaries and with coincident $x$-- and $x'$-- axes. Therefore, the
relative velocity of the two adjacent grains is 

\begin{equation}
\begin{aligned}
&\Delta v_{i'}=
\begin{cases}
\mathcal{Q}\,\sigma_{i'z'}
\left( 1-\displaystyle\frac{\tau_c}{\tau_{z'}}\right),\,
&i'=x',y',\,\text{if }\,\tau_{z'}>\tau_c \\
0, &\text{ otherwise},  
\end{cases} \\
&\Delta v_{z'}\equiv 0,,
\label{E2}
\end{aligned}
\end{equation}

\noindent
where $\mathcal{Q}$ is a coefficient depending only on the local pressure
$p$ and temperature $T$. By the procedure explained in Ref.~\cite{Lagos1}
the force law (\ref{E2}) can be written in terms of the strain rates
$\dot{\varepsilon}_{ij}$ instead of the relative velocities $\Delta\vec{v}$
between adjacent grains, and expressed in the common $(x,y,z)$ frame of
reference. The procedure gives

\begin{equation}
\begin{aligned}
&\frac{\dot\varepsilon_{zz}}{\sigma +p}
=\frac{3\mathcal{Q}}{8d}f_s(2\theta_c), \\
&\frac{\dot\varepsilon_{xx}}{\sigma +p}
=\frac{\dot\varepsilon_{yy}}{\sigma +p}
= \frac{3\mathcal{Q}}{8d}g_s(2\theta_c),
\end{aligned}
\label{E3}
\end{equation}

\noindent
where $\theta_c$ is defined by

\begin{equation}
\sin(2\theta_c)=\frac{4\tau_c}{3|\sigma+p|},
\label{E4}
\end{equation}

\noindent
and $f_s(2\theta_c)=\cos (2\theta_c)-s(\pi/2-2\theta_c)\sin(2\theta_c)$, $g_s(2\theta_c)=1-4\theta_c/\pi +(1/\pi)(1-2s)\sin(4\theta_c)$.

\begin{figure}[h!]
\begin{center}
\includegraphics[width=8cm]{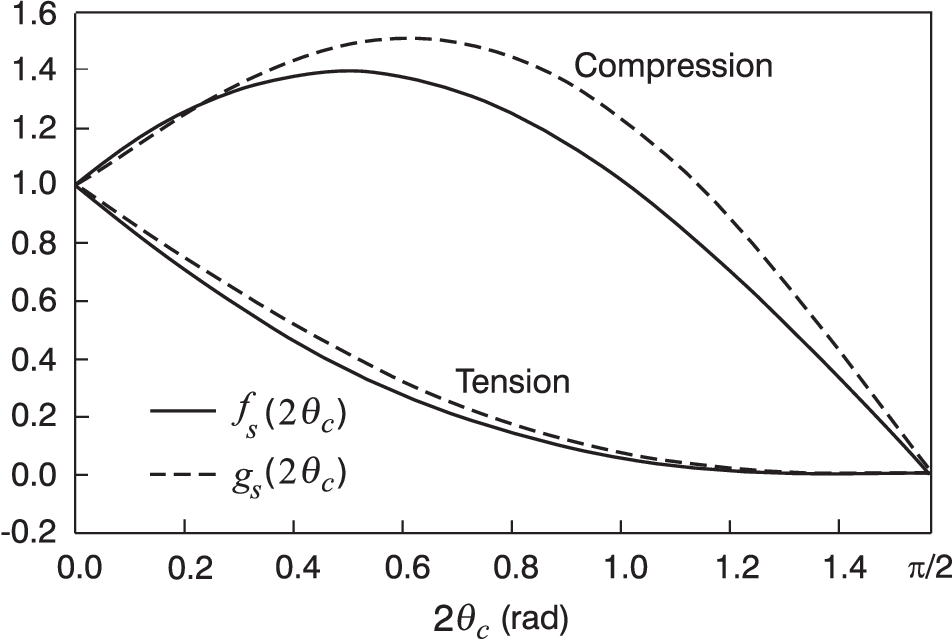}
\caption{\label{Fig1} Functions $f_s$ and $g_s$. L\'evy--Mises
Eqs.~(\ref{E6}) are exact when $f_s(2\theta_c)=g_s(2\theta_c)$.}
\end{center}
\end{figure}

Functions $f_s$ and $g_s$, shown in Fig.~\ref{Fig1}, have real values for
$|\sigma+p|\ge 4\tau_c/3$, which constitutes the yield condition. By the
cylindric symmetry, this condition can also be written as

\begin{equation}
(\sigma_x+p)^2+(\sigma_y+p)^2+(\sigma_z+p)^2
\ge \frac{8}{3}\tau_c^2 ,
\label{E5}
\end{equation}

\noindent
which has the same general form of the von Mises yield criterion, first
formulated by Huber \cite{Hill, Khan}. As well, Eqs.~(\ref{E3}) reduce to
the L\'evy--Mises equations \cite{Hill, Khan} 

\begin{equation}
\frac{\dot\varepsilon_{xx}}{\sigma_x+p}
=\frac{\dot\varepsilon_{yy}}{\sigma_y+p}
=\frac{\dot\varepsilon_{zz}}{\sigma_z+p},
\quad\text{(L\'evy--Mises)}
\label{E6}
\end{equation}

\noindent
for the plastic flow of solid materials when $\theta_c\approx 0$. In
previous papers we have demonstrated that plastic and superplastic flow
takes place whenever

\begin{equation}
\begin{aligned}
\theta_c\phantom{a} 
\begin{cases}
&\gtrapprox\, 0, \quad\text{plastic flow}, \\
&\lessapprox\,\pi/4, \quad\text{superplastic flow}.
\end{cases}
\end{aligned}
\label{E7}
\end{equation}

The coefficient $\mathcal{Q}(p,T)$ is given by

\begin{equation}
\frac{\mathcal{Q}}{4d}=C_0\frac{\Omega^*}{k_BT}
\exp\left( -\frac{\epsilon_0+\Omega^*p}{k_BT}\right)
\label{E8},
\end{equation}
 
\noindent
where $k_B$ is the Boltzmann constant, $C_0$ is a coefficient depending
only on the mean grain size $d$, the constant $\epsilon_0$ is the energy
necessary for evaporating a crystal vacancy from the grain boundary to the
crystalline grain, and $\Omega^*$ the local volume variation in the same
process \cite{Lagos1,LagosDuque,Vetrano}. On the other hand, by Hooke's
law $B(\Delta V/V)=-p$, where $B$ is the bulk modulus, and
$\dot{V}=\dot{\varepsilon}_{xx}+\dot{\varepsilon}_{yy}+
\dot{\varepsilon}_{zz}$, we have that

\begin{equation}
\dot{p}=B(\dot{\varepsilon}_{xx}+\dot{\varepsilon}_{yy}+
\dot{\varepsilon}_{zz}).
\label{E9}
\end{equation}

Combining Eqs.~(\ref{E3}), (\ref{E4}), (\ref{E8}) and (\ref{E9}),
retaining up to first order terms in $\theta_c$ to settle the condition
for plastic deformation, and proceeding as explained in Refs.~\cite{Lagos3}
and \cite{Lagos4}, one arrives to

\begin{equation}
\sigma =\sigma_0+C\varepsilon +D\ln\bigg( 1-\frac{|\varepsilon|}{\varepsilon_{\text{frac}}}\bigg),
\label{E10}
\end{equation}

\noindent
where $\sigma_0$ is the yield stress, and the other constants are

\begin{equation}
C=\frac{2}{3}\bigg(\pi -\frac{8}{\pi}\bigg)\frac{B|\Omega^*|\tau_c}{k_BT},
\quad
D=\frac{k_BT}{|\Omega^*|}
\label{E11},
\end{equation}

\begin{equation}
\varepsilon_{\text{frac}}=
\frac{\pi}{(\pi^2-8)}\frac{k_B T}{B\Omega^*}
\frac{\sigma_0}{\tau_c}.
\noindent
\label{E12}
\end{equation}

\begin{figure}
\begin{center}
\includegraphics[width=8cm]{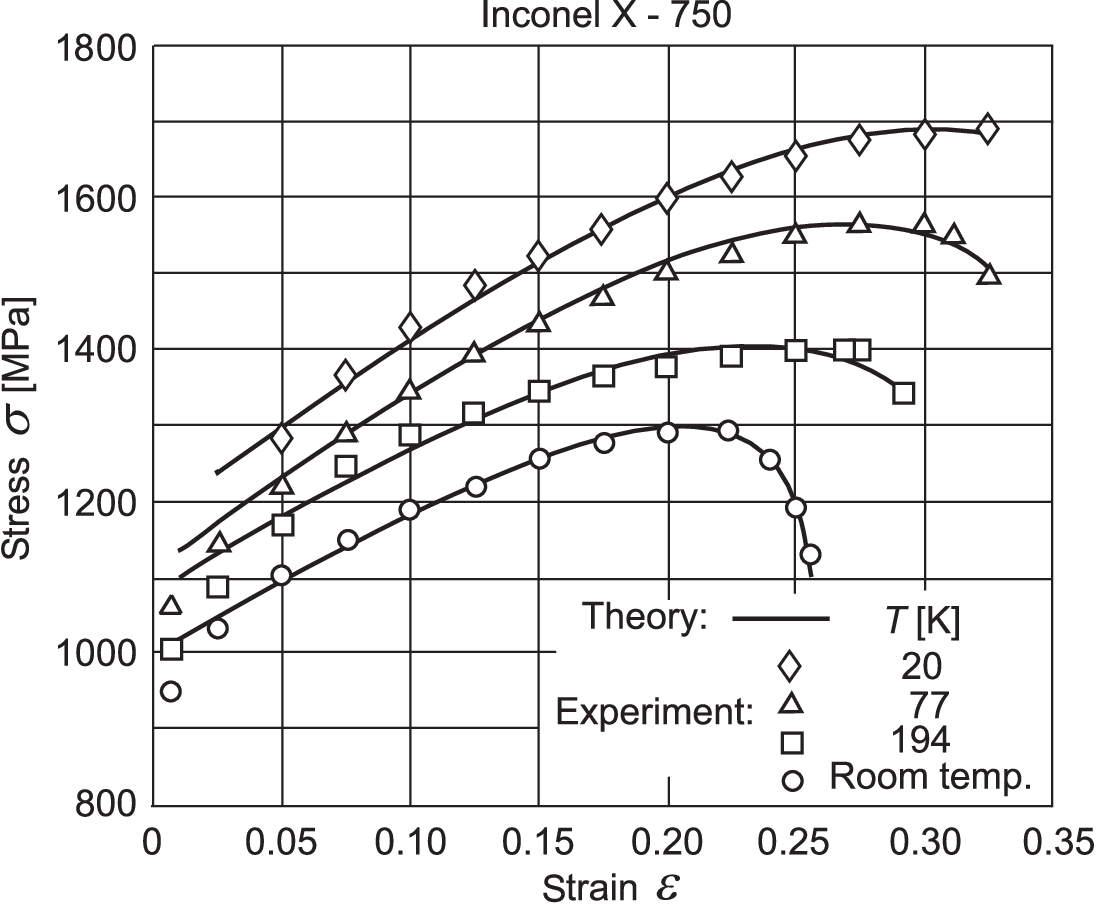}
\caption{\label{Fig2} Discrete symbols represent experimental data for the
stress--strain rate for the nickel superalloy Inconel X750 at room and lower
temperatures \cite{Wolf}. Solid lines are the predictions of Eq.~(\ref{E10})
with the parameters of Table {\ref{table1}} and Fig.~\ref{Fig3}.}
\end{center}
\end{figure}

\begin{center}
\begin{table} 
\caption{\label{table1} Values of the parameters for Eq.~(\ref{E10}) giving
the fits of Fig.~\ref{Fig2}.}
\begin{center}
\begin{tabular}{ccccc}
Parameter & 20K & 77K & 194K & 293K \\
\hline
$\sigma_0$ [MPa] & 1175 & 1112 & 1080 & 1000 \\
$C$ [MPa] & 3740 & 3720 & 3400 & 2480 \\
$D$ [MPa] & 550 & 540 & 475 & 135 \\
$\varepsilon_{\text{frac}}$ & 0.448 & 0.420 & 0.373 & 0.261 \\
\hline
\end{tabular}
\end{center}
\end{table}
\end{center}

\begin{figure}[h!]
\begin{center}
\includegraphics[width=7cm]{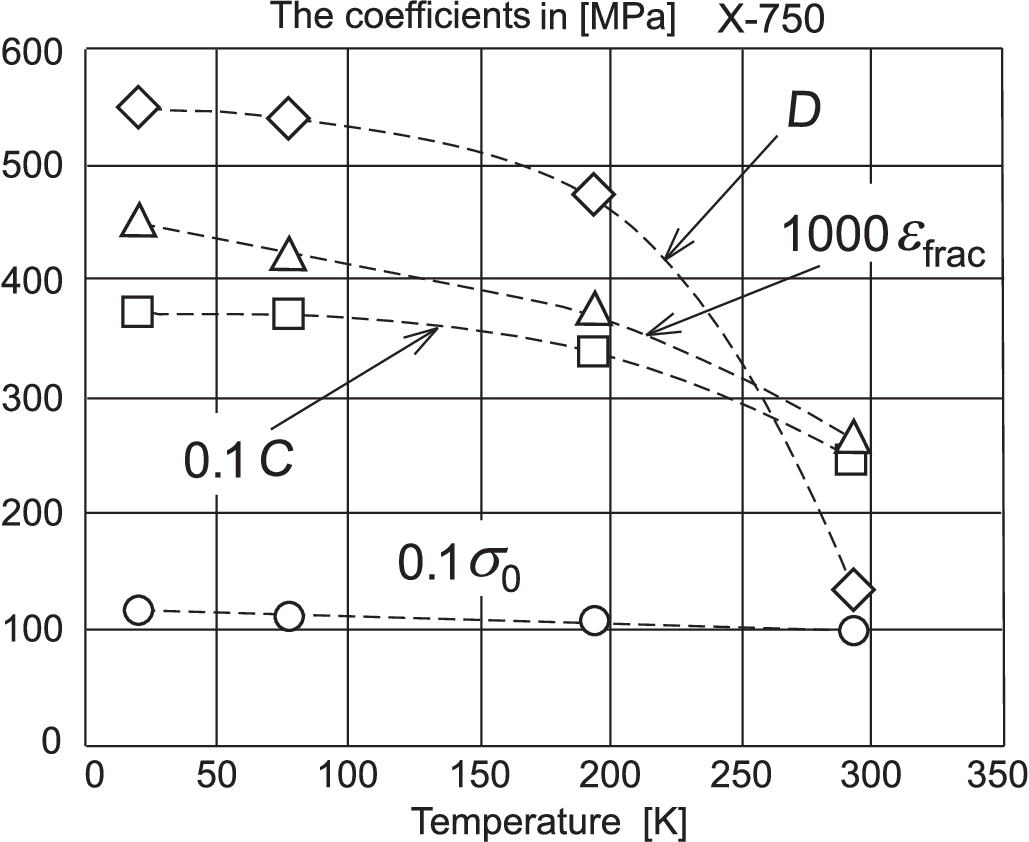}
\caption{\label{Fig3} Graphical display of the constants in Table
\ref{table1} showing their dependence on temperature.}
\end{center}
\end{figure}

Fig.~\ref{Fig2} shows a comparison of the experimental data
\cite{Wolf} on the nickel superalloy Inconel X-750 at room and lower
temperatures with the predictions of Eq.~(\ref{E10}). The discrete symbols
in Fig.~\ref{Fig2} represent the experimental data of the true strains
$\varepsilon$ that Inconel X-750 bars undergo when subjected to axial true
stresses $\sigma$ \cite{Wolf}, at room and lower temperatures. The
continuous lines are obtained from Eq.~(\ref{E10}) with the constants of
Table \ref{table1} for the four temperatures. The values for the constants
appearing in Table \ref{table1} are graphically displayed in Fig.~\ref{Fig3}
to show their dependence on the temperature $T$. 

\begin{figure}[h!]
\begin{center}
\includegraphics[width=7cm]{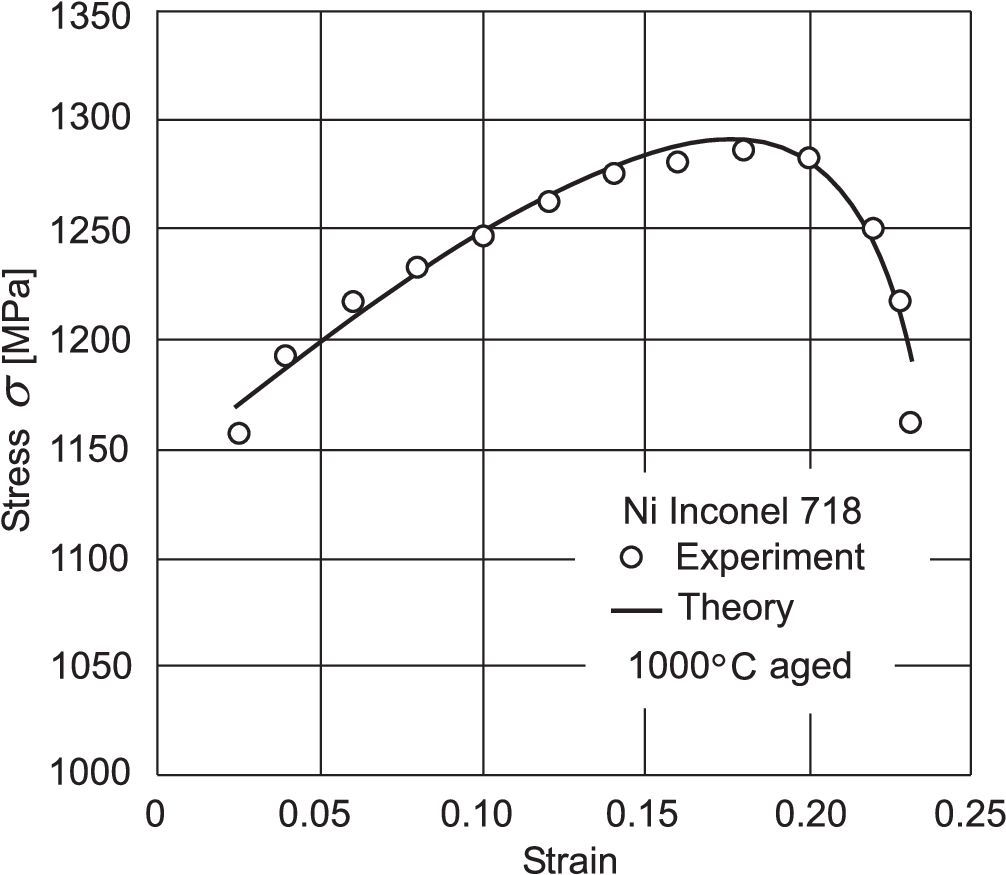}
\caption{\label{Fig4} Tensile stress-strain behaviour of additively
manufactured Inconel 718 bars after one hour heat treatment and ageing.}
\end{center}
\end{figure}

\begin{figure}[h!]
\begin{center}
\includegraphics[width=7cm]{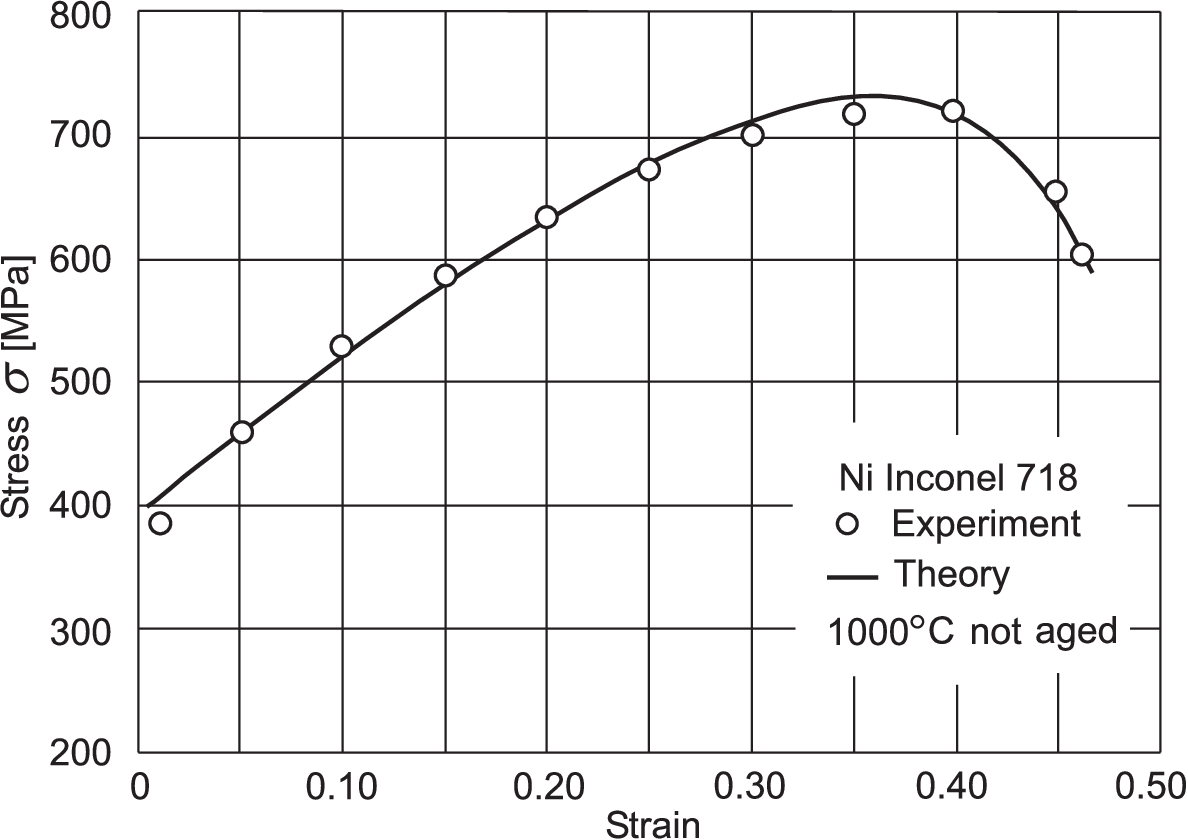}
\caption{\label{Fig5} Same as Fig.~\ref{Fig4}, but previous to the ageing
procedure.}
\end{center}
\end{figure}

Figs.~\ref{Fig4} and \ref{Fig5} depict the measured tensile stress--strain
curves of additively manufactured bars of the nickel superalloy Inconel
718 \cite{Brennel}, and the curves given by Eq.~(\ref{E10}) with the
parameters of Table \ref{table2}. The big difference exhibited by the two
graphs shows that the true stress--strain analysis constitutes a good
method for determining the optimal condition of a given material.

\begin{center}
\begin{table}[h!]
\caption{\label{table2} Values of the parameters for Eq.~(\ref{E10}) giving
the fits of Figs.~\ref{Fig4} and \ref{Fig5}.}
\begin{center}
\begin{tabular}{ccccc}
Parameter & 1000K not aged & 1000K aged \\
\hline
$\sigma_0$ [MPa] & 395 & 1140 \\
$C$ [MPa] & 1860 & 1715 \\
$D$ [MPa] & 270 & 118 \\
$\varepsilon_{\text{frac}}$ & 0.507 & 0.244 \\
\hline
\end{tabular}
\end{center}
\end{table}
\end{center}

\begin{figure}[h!]
\begin{center}
\includegraphics[width=8cm]{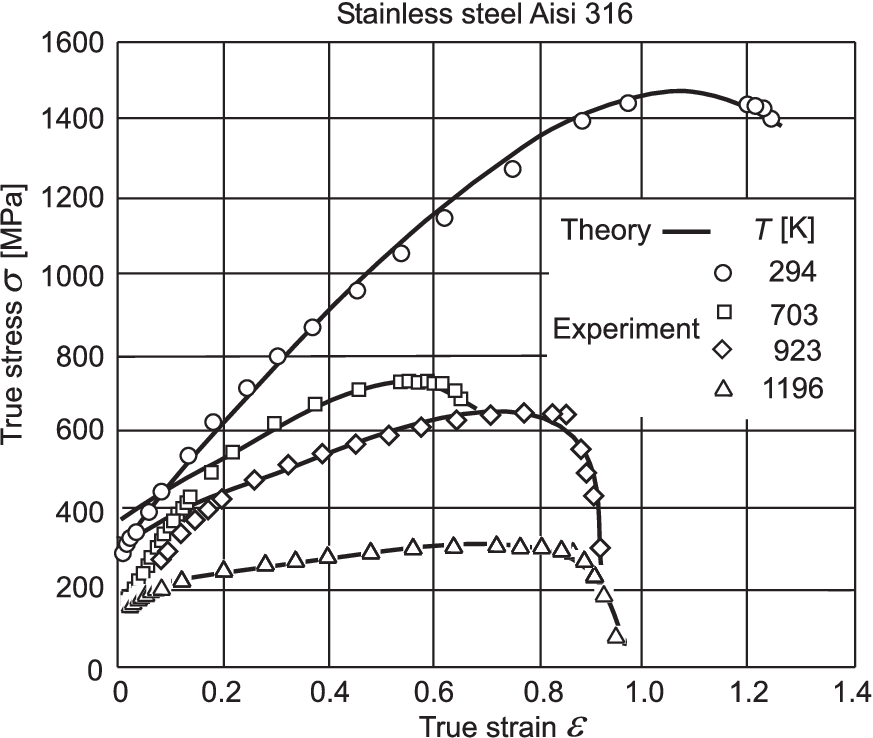}
\caption{\label{Fig6} True strain--true stress data for the stainless steel
AISI 316 tested at room and high temperatures. Discrete symbols represent
the experimentally observed data and solid lines Eq.~(\ref{E10}) with the
parameters of Table \ref{table3}.}
\end{center}
\end{figure}

Fig.~\ref{Fig6} and Table \ref{table3} show the experimentally measured
true stress--strain at room and much higher temperatures of the stainless
steel AISI 316 \cite{Conway}. The stresses and corresponding strains were
determined with simultaneous control of the uniform diameter of the steel
probes, in order to obtain the true values. No necking is observed when
the stress reaches a maximum, indicating a true material weakening before
fracture. One can interpret this saying that fracture is not a direct
consequence of the applied external force, but mainly of the pressure $p$
cumulated during the plastic deformation. In effect, $p$ is an important
variable which cannot be ignored. Ignoring $p$ may produce a deceivingly
inconsistent behaviour of the material in different situations, frequently
attributed to a vague dependence on history. Most times this dependence is
simply determined by a different value of the variable $p$. When the
material is subjected to a series of mechanical tests the value reached by
$p$ in one of them is the initial condition for the next one.

The different behaviour for tensile and compressive plastic events
exhibited in Fig.~\ref{Fig1} may have strong effects in the plastic
response of the material when subjected to non uniform stresses.
Bauschinger effect, fatigue, creep and other properties will be dealt with
in a future communication.

\begin{center}
\begin{table} [h!]
\caption{\label{table3} Values of the parameters for Eq.~(\ref{E10})
giving the fits of Fig.~\ref{Fig6}.}
\begin{center}
\begin{tabular}{ccccc}
Parameter & 293K & 703K & 923K & 1196K \\
\hline
$\sigma_0$ [MPa] & 310 & 374 & 330 & 198 \\
$C$ [MPa] & 2600 & 1070 & 745 & 336 \\
$D$ [MPa] & 1623 & 139 & 140 & 88 \\
$\varepsilon_{\text{frac}}$ & 1.695 & 0.690 & 0.925 & 0.958 \\
\hline
\end{tabular}
\end{center}
\end{table}
\end{center}

\end{document}